%% file: main.tex
\documentclass[11pt,a4paper]{article}
\pdfoutput=1

\usepackage{commands}

\usepackage{pst-grad} 
\usepackage{xcolor}

\usepackage{fullpage}
\newenvironment{proof}{Proof: }{}
\newcommand{\affaddr}[1]{\small{#1}}
\newcommand{\email}[1]{\small{#1}}
\newcommand{\numberofauthors}[1]{}
\newcommand{\alignauthor}{\\\ \\}

\title{Continuous Top-k Queries over Real-Time Web Streams
\longversion{\linebreak(extended version)}}

\numberofauthors{3}

\author{
\alignauthor
Nelly Vouzoukidou\\
\affaddr{Sorbonne Universités, UPMC Univ Paris 06, UMR 7606, LIP6}\\
       \affaddr{Paris, France}\\
       \email{nelly.vouzoukidou@lip6.fr}
\alignauthor
Bernd Amann\\
\affaddr{Sorbonne Universités, UPMC Univ Paris 06, UMR 7606, LIP6}\\
       \affaddr{Paris, France}\\
       \email{bernd.amann@lip6.fr}
\alignauthor
Vassilis Christophides\\
       \affaddr{INRIA Paris, LINCS}\\
       \affaddr{Paris, France}\\
       \email{vassilis.christophides@inria.fr}
}

\renewcommand{\rem}[3]{}
\renewcommand{\longversion}[1]{}
\renewcommand{\withoutproofs}[1]{}
\begin{document}

\maketitle

\begin{abstract}
  The Web has become a large-scale real-time information system forcing us to revise both how to effectively assess relevance of information for a user and how to efficiently implement information retrieval and dissemination functionality. To increase information relevance, Real-time Web applications such as Twitter and Facebook, extend content and social-graph relevance scores with  ``real-time'' user generated events (e.g.\ re-tweets, replies, likes). To accommodate high arrival rates of information items and user events we explore a publish/subscribe paradigm in which we index queries and update on the fly their results each time a new item and relevant events arrive. In this setting, we need to process \emph{continuous top-$k$ text queries} combining both \emph{static} and \emph{dynamic} scores. To the best of our knowledge, this is the first work addressing how \emph{non-predictable}, dynamic scores can be handled in a continuous top-$k$ query setting.
\end{abstract}






\input{introduction}

\input{problemStatement}

\input{eventHandler}

\input{candidateIndexing}

\input{experiments}

\vspace{-2mm}
\input{relatedWork}

\input{conclusions}

\bibliographystyle{abbrv}
\bibliography{biblio}
\end{document}

%% file: introduction.tex
\section{Introduction}

The proliferation of social media platforms and mobile technology enables people to almost instantly produce and consume information world-wide. In the \emph{real-time Web}, traditional scoring functions based on content similarity or link graph centrality are no longer sufficient to assess \emph{relevance} of information to user needs. For example, textual \emph{content} relevance may severely be obscured by the ``churn'' observed in the term distribution of real-time media news or social content and those found in historical query logs~\cite{DBLP:conf/icwsm/LinM12}, while \emph{social relevance}~\cite{DBLP:conf/www/KhodaeiS12} of short living information cannot be reliably computed on static graphs. Streams of user events like ``replies'' (for posting comments), ``likes'' (for rating content) or ``retweets'' (for diffusing information) represent nowdays valuable social feedback~\cite{Kwak:2010:TSN:1772690.1772751} on web content\footnote{This trend is expected to be amplified in the future Internet of Things (IoT) where users can provide online feedback regarding almost any digital or physical object~\cite{CP15}.} that is not yet exploited online to assess \emph{contextual} relevance aspects. 

\emph{Real-time Web search engines} are essentially confronted with a double challenge. First, ``ephemeral'' Web content~\cite{DBLP:journals/pvldb/GrinevGHK11} should become searchable for millions of users immediately after being published. Second, \emph{dynamic} relevance of streaming information (e.g.\ \emph{user attention}~\cite{WZRM08,LDP10}, \emph{post credibility}~\cite{Gupta:2012:CRT:2185354.2185356}, \emph{information recency}~\cite{Dong:2010:TEI:1772690.1772725}) should be continuously assessed as users provide their feedback.

To tackle both content and score dynamicity, a first solution consists in extending a content retrieval model (textual, spatial) and implementing adequate index structures and refresh strategies for \emph{low-latency} and \emph{high throughput snapshot query evaluation}. Best-effort refresh strategies~\cite{cho03,HAA12} could determine optimal re-evaluation of active user queries which combined with real-time content indexing~\cite{CLOW11,BGL+12,WLXX13,MME+14,LBLT15} can achieve high result freshness and completeness. However, real-time content indexing systems usually accommodate high arrival rates of items at the expense of result accuracy by either (a) excluding a significant portion of the incoming items (e.g.\ with infrequent keywords) from the index to reduce update costs or (b) by ranking items only by arrival time to support append-only index insertion and thus ignore content relevance to queries~\cite{WLXX13}. Coupling both \emph{static} (e.g.\ content similarity to user queries) with \emph{dynamic} aspects of relevance beyond information recency \cite{Dong:2010:TEI:1772690.1772725} is clearly an open issue in existing real-time search engines. 

An alternative solution is a publish/subscribe architecture~\cite{HMA10,MP11,VAC12, Shraer:2013:TPS:2536336.2536340, Chen:2013:EQI:2463676.2465328, Guo:2015:LPS:2723372.2746481, DBLP:conf/icde/ChenCCT15} in which user 
\emph{(continuous) queries} rather than information items are indexed while their results are updated on the fly as new items arrive. In \emph{predicate-based} \cite{Chen:2013:EQI:2463676.2465328,Guo:2015:LPS:2723372.2746481} publish/subscribe systems, incoming items that match the filtering predicates are simply added to the result list of continuous queries, while in \emph{similarity-based top-k} \cite{HMA10,MP11,VAC12,Shraer:2013:TPS:2536336.2536340, DBLP:conf/icde/ChenCCT15} systems, matching items have also to exhibit better relevance w.r.t. the items already appearing as the top-k results of continuous queries. In publish/subscribe systems both (a) \emph{early pruning of the query index traversal} for locating relevant queries to an incoming item and (b) the \emph{efficient maintenance of their results lists} are challenging problems. Only recently, these problems have been studied for more dynamic settings such as decaying information relevance as time passes (e.g.\ for textual~\cite{Shraer:2013:TPS:2536336.2536340,VAC12} or spatio-textual~\cite{DBLP:conf/icde/ChenCCT15} streams). However, there is an important difference between information decay and social feedback: while information freshness is defined as a \emph{known in advance, global} function applied on all item scores simultaneously, social feedback is \emph{locally} defined as a stream of non-predictable, random events per item. First, such randomness introduces a new degree of freedom requiring adaptive optimization strategies to the incoming feedback events. Second, the item score dynamics significantly varies and calls for new techniques to balance processing cost (memory/CPU) between items with high and low feedback.

In this paper we are studying a new class of continuous queries featuring \emph{real-time scoring functions} under the form of \emph{time decaying positive user feedback} for millions of social feedback events per minute and millions of user queries. In a nutshell, the main contributions of our work are:
\begin{itemize}
\item We introduce continuous top-$k$ queries over real-time web
  streams aggregating both \emph{static item content} and
  \emph{dynamic real-time event} scores. Then, we decompose the
  continuous top-$k$ query processing problem into \emph{two separate
  tasks} matching item contents and feedback events.
\item We propose new \emph{efficient in-memory data structures} for indexing continuous top-$k$ queries when matching incoming user events along with a family of \emph{adaptive algorithms for maintaining the result of top-$k$ queries with highly dynamic scores}.
\item We experimentally evaluate our index structures over a real-world dataset of 23 million tweets collected during a 5-month period and compare the impact of dynamic scores in the performance of our algorithms and associated data structures.
\end{itemize}
The rest of the paper is organized as follows. Section~\ref{sec:dyn_problemStatement} gives a formal definition of the problem. Section~\ref{sec:dyn_eventHandler} presents our proposed solution and pruning techniques for event matching. Then, Section~\ref{sec:dyn_candIndexing} describes a number of index implementations that apply these techniques whose experimental evaluation is detailed in Section~\ref{sec:dyn_experiments}. Related work is presented in Section~\ref{sec:relatedWork} and the main conclusions drawn from our work are summarized in Section~\ref{sec:dyn_discussion}.

%% file: problemStatement.tex
\section{Real-time top-$k$ queries}
\label{sec:dyn_problemStatement}
In this section we formally define the problem of evaluating continuous top-$k$ queries with dynamic scores.

\subsection{Data Model}
The general data model builds on a set of search queries $\queries$, a
set of items $\items$ and a set of events $\events$.  \ber{do we need
  the following constraint?:} Each feedback event $\ev \in \events$ (e.g.,\ ``replies'', ``likes'', ``retweets'') concerns exactly one target item $\ite$ denoted by $\target(\ev)$ (e.g.,\ media news articles, social posts). The set of all events with target $\ite$ is denoted by \itemevents{\ite}. We also assume a time-stamping function $\ts: \queries \cup \items \cup
\events \rightarrow \T$ which annotates each query, item and event
with their arrival time-stamp in the system. This function introduces
a weak ordering between queries, items and events and formally
transforms all sets $\queries$, $\items$ and $\events$ into streams
\querystream{},\ \itemstream{} and \eventstream{}.

\paragraph{Continuous top-k queries}
A continuous top-$k$ query $\qu=(k, \queryscorefct, \aggeventscorefct,
\alpha)$ is defined by a positive constant $k$, a static query score function $\queryscorefct$, an event score function $\eventscorefct$ and some non negative query score weight $\alpha\leq 1$:

\begin{itemize}
\item The constant $k$ defines the maximum \emph{size} of the query result at each time instance.
\item The \emph{query score} function $\queryscorefct : \queries\times \items \rightarrow [0,1]$ returns a static score of query $\qu\in\queries$ and item $\ite\in\items$.  
It may capture popular content similarity measures (e.g.\ textual like cosine, Okapi BM25 or spatial similarity), but it might also reflect other static, query independent scores like source authority~\cite{MLYZG10} or media focus~\cite{MC10}. 
\item The \emph{event score} function $\eventscorefct: \events \rightarrow [0,1]$ returns a positive score for each $\ev\in\events$. Events can be scored according to their \emph{type and content} (e.g.\ comment, share, click, like)~\cite{Duan10} as well as their \emph{author} (e.g.\ friend versus anonymous~\cite{AVB15a}). Distance measures between the author of an even the user subscribing a query could be also considered~\cite{LBLT15}.  
\item The \emph{dynamic (feedback) score} function $\aggeventscorefct  : \items \times \T  \rightarrow  [0,\infty]$ aggregates the scores of all  events $\ev \in \itemevents{\ite}$ with target $\ite$ up to time instant $\ti$:
  \begin{equation*}
    \aggeventscore{\ite}{\ti} = \sum_{\ev \in \itemevents{\ite}, ts(\ite) \leq ts(\ev) \leq \ti} \eventscore{\ev}
    \label{eq:aggeventscore}
  \end{equation*}
  Observe that we only allow positive event scores and the aggregated
  event score is unbound.  \ber{is unboundedness true?}
\item The \emph{total score} of some item \ite{} w.r.t.\ some query
  \qu{} is defined a linear combination of the query score
  \queryscore{\qu}{\ite}: 
  \begin{eqnarray*}
\sTotal(\qu, \ite, \ti)=\alpha \cdot
  \queryscore{\qu}{\ite} + (1-\alpha) \cdot
  \aggeventscore{\ite}{\ti}
\label{eq:totalscore}
\end{eqnarray*}
Observe that feedback on items is ignored when $\alpha=1$ whereas
items are only ranked with respect to their feedback (independently
from the query score) when $\alpha=0$. As a matter of fact, we consider a general form of \emph{static} and \emph{dynamic} scoring function that abstracts several score aspects of items proposed in the literature.
\end{itemize}

\begin{example}
\begin{figure}[htbp]
  \centering
\includegraphics[width=\textwidth]{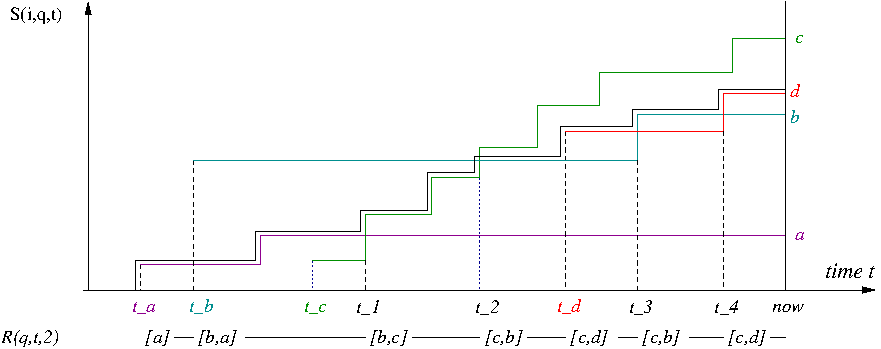}
  \caption{Top-2 result evolution of query $q$}
  \label{fig:top-2}
\end{figure}
Figure~\ref{fig:top-2} illustrates the high dynamicity of the query
result of a single continuous top-$k$ query over real-time web streams.
It shows the evolution of the top-$2$ result
of some $\qu$ for item set $\itemstream{now}=\{a,b,c,d\}$. The score
evolution of each item is represented by a stepwise line where each
monotonic increase corresponds to the arrival of some event
(aggregated event score). The minimum score of \qu\ is represented by
a bold line. Each item has an initial positive query score and the
query result is updated seven times: at the arrival of a new item
(item match at $\tau_a$, $\tau_b$ and $\tau_d$) as well as at the
arrival of a new event (event match at $\tau_1$, $\tau_2$, $\tau_3$
and $\tau_4$). Observe that the arrival of a new item does not
necessarily trigger a query result update (e.g. arrival of $c$) and an
item can disappear and reappear in the result (item $b$ disappears at
$\tau_d$ and reappears at $\tau_3$).
\end{example}

\paragraph{Decay function}
To take account of information
freshness~\cite{Dong:2010:TEI:1772690.1772725}, we consider an
order-preserving decay function $\decay{s}{d}$ which can be applied to
query, item or event score values $s$: given a time duration $d$ such
that $\decay{s}{0}=s$ and $\decay{s}{d}\leq \decay{s}{d'}$ for all
$d>d'$, i.e.\ if $s>s'$, then $\decay{s}{d}>\decay{s'}{d}$ for any
duration $d$. Order-preserving decay functions\footnote{Whereas any
  linear or exponential function is order-preserving, it is possible
  to define polynomial functions which are not order preserving.}
allow in particular the use of the backward decay
techniques~\cite{CSSX09} where all scores are computed, stored and
compared with respect to some \emph{fixed reference time instant}
$\tau_0$ used as a landmark.  It should be stressed that two possible
semantics for event decay can be defined. The \emph{aggregated decay
  semantics} consists in applying decay to the aggregated event score
with respect to the target item age, whereas the \emph{event decay
  semantics} consists in applying decay to each individual event with
respect to the event age. In the first case, all events have the same
decay, whereas the second case favors recent events to old ones.  In
the special case of linear decay both semantics lead to the same
results. The aggregated decay semantics maintains the order preserving
property, whereas the event decay semantics might change the item
order for order preserving exponential decay.  In the following and in
our experiments, we assume a linear order-preserving decay function.

\paragraph{Formal semantics and problem statement}

We denote by \ritems{\qu}{\ti} the set of \emph{relevant} items which
have a strictly positive query score $\queryscore{\qu}{\ite}>0$. The
\emph{top-$k$ result} of a continuous query $\qu=(k, \queryscorefct,
\aggeventscorefct, \alpha)$ at some time instant $\ti$, denoted
$\topk{\qu}{\ti}$, contains the subset of maximally $k$
\emph{relevant} items $\ite\in\ritems{\qu}{\ti}$ with the highest
total scores $\sTotal(\qu, \ite, \ti)$. In other words, for each item
$\ite\in \topk{\qu}{\ti}$ there exists no other (relevant) item $\ite'
\in \items-\topk{\qu}{\ti}$ with a higher total score $\sTotal(\qu,
\ite', \ti)>\sTotal(\qu, \ite, \ti)$. Using the backward decay
technique described before, we suppose that all scores are computed, stored and compared with respect to some \emph{fixed reference time   instant} $\tau_0$ used as a landmark.

We can now state the general real-time top-$k$ query evaluation problem which we will address in the rest of this paper:

\sloppy
\begin{problem}
Given a query stream $\queries^\ts$, an item stream $\items^\ts$, an
event stream $\events^\ts$ and a total score function
$\globalscorefct$ (with order-preserving decay), maintain for each
query $\qu\in \queries^\ts$ its continuous top-$k$ result
$\topk{\qu}{\ti}$ at any time instant $\ti\geq \ts(q)$.
\label{prob:querymatch}
\end{problem}


\subsection{Query execution model}
This top-k query maintenance problem can be reformulated into a sequence of updates triggered by the arrival of new queries, items and events at
different time-instants. We will consider a fixed time instant $\ti$ and
denote by
\begin{itemize}
\item $\pqueries{\ite}{\ti}=\{\qu|\ite \in \topk{\qu}{\ti}\}$, the
  set of \emph{active queries}  where item $\ite$ is published at time instant $\ti$;
\item $\citems{\qu}{\ti}=\{\ite| \ite \not\in \topk{\qu}{\ti} \wedge \queryscore{\qu}{\ite} > 0\}$, the set of all \emph{candidate items} which might be inserted into the result of $\qu$ on a later time instant $\ti'>\ti$ due to feedback score updates.
\item $\cqueries{\ite}{\ti}=\{\qu|\ite \in \citems{\qu}{\ti}\}$, the set of all \emph{candidate queries} for item \ite;
\end{itemize}
%


\begin{example}
\label{ex:update}
For the example of Figure~\ref{fig:top-2}, during time interval
$(\tau_2,\tau_d]$ , query $\qu$ appears in the active sets of items
$c$ and $b$. After the arrival of item $d$, query $\qu$ is inserted into
the active set of $d$ and moves from the active set to the candidate
set of $b$. The arrival of a new event targeting $b$ at time instant
$\tau_3$ switches $\qu$ from the active (candidate) set to the
candidate (active) set of $d$ ($b$).
\end{example}

Problem statement~\ref{prob:querymatch} can be decomposed into three separate matching tasks at a given time instant $\ti$:


\begin{description}
\item[Query matching:]
Given a query $\qu$, compute the set $\topk{\qu}{\ti}^+$ of items that should be added to the top-$k$ result of $\qu$;

\item[Item matching:]
\label{prob:itemmatch}
Given an item $\ite$, identify the set
$\pqueries{\ite}{\ti}^+$
 of all queries that should be updated by
adding $\ite$ in their result.


\item[Event matching:]
\label{prob:eventmatchadd}
Given an event $\ev$ with target item $\ite=target(\ev)$, identify
the set $\pqueries{\ite}{\ti}^+$ of all
queries that should be updated by adding $\ite$ to their result.
\end{description}


Observe that through the definitions of \pqueries{\ite}{\ti} and
\topk{\qu}{\ti} all three matching tasks are strongly related and some
tasks could be partially solved by using others. For example, item
matching can be solved by re-evaluating (refreshing the result of) all
queries while event matching can be solved by re-executing the item
matching task for its target item with the updated event score.  We
will show in Section~\ref{sec:dyn_eventHandler} and experimentally verify in in Section~\ref{sec:dyn_experiments}, this solution exhibits serious performance limitations for information items with highly dynamic scores.  In the rest of this article we are mainly interested in the definition and implementation of an efficient solution for the event matching task by relying on existing efficient solutions for the query and item matching task.


\paragraph{Query evaluation architecture and algorithm}

\begin{figure}[htbp]
\center
\includegraphics[width=0.4\linewidth]{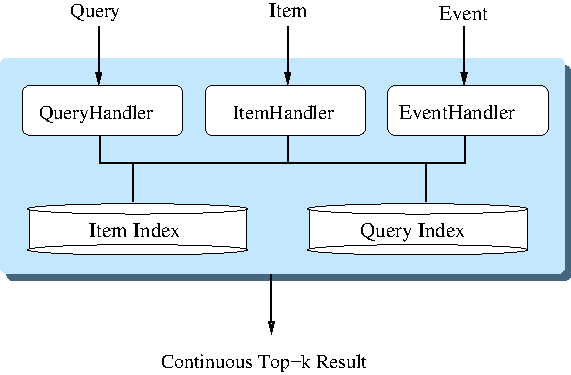}
\caption{Architecture}
\label{fig:intro_architectureAll}
\end{figure}

As a matter of fact, by considering distinct matching tasks, we can
devise a modular architecture and experiment with different data
structures and algorithms. As depicted in
Figure~\ref{fig:intro_architectureAll} the main processing modules
\QH{} (\QHshort), \IH{} (\IHshort) and \EH{} (\EHshort) are
independent and share data structures for indexing queries (\QI) and
items (\II). Events are not stored but dynamically aggregated in the
corresponding item scores and updated in the \II.

\SetKwProg{Proc}{}{}{} \SetKwProg{Fn}{}{}{}
\SetKwFunction{KwprocessQuery}{QH.processQuery}
\SetKwFunction{KwprocessEvent}{EH.processEvent}
\SetKwFunction{KwprocessItem}{IH.processItem}
\SetKwFunction{KwmatchEvent}{EH.matchEvent}
\SetKwFunction{KwindexItem}{II.addItem}
\SetKwFunction{KwupdateItem}{II.updateItem}
\SetKwFunction{KwindexQuery}{QI.addQuery}
\SetKwFunction{KwmatchItem}{QI.matchItem}
\SetKwFunction{KwmatchQuery}{II.matchQuery}
\SetKwFunction{Kwadd}{RTS.add}
\SetKwFunction{Kwdel}{RTS.del}
\SetKwFunction{KwmatchEventSimple}{EH.matchEventSimple}

The general interaction of these models for processing incoming
queries, events and items is illustrated in
Algorithm~\ref{alg:processing}.
 Function \KwprocessQuery{} indexes each new incoming
  query $\qu$.
  Functions \KwprocessItem{} and \KwprocessEvent{} match new incoming items and events
  against the registered queries and identify the queries whose
  result (top-$k$ set) have to be updated.
  Function \KwmatchItem computes $\pqueries{\ite}{\ti}^+$ for a given
  item (contents) $\ite$ with a positive constant event score. This
  supposes that the \IH\ supports continuous top-$k$ queries with
  \emph{inhomogeneous} score functions aggregating a
  \emph{query-dependent score} (e.g. cosine similarity) and a
  \emph{query-independent} constant value (for example item popularity).   
  Such an index structure has been proposed for example in~\cite{VAC12}.
  \ber{est-ce qu'il y a d'autres exemples}

  Function \KwmatchEvent{} matches each new incoming event and
  generates $\pqueries{target(\ev)}{\ti}^+$. In the rest of this
  paper we are mainly interested in the efficient implementation of
  this function and our approach will be described in
  Section~\ref{sec:dyn_eventHandler}.
Function \Kwadd{} publishes item $\ite$ in the result of
  $\qu$. It also updates the publication and candidate sets for
  $\ite$ and query $\qu$ and, if necessary, for item $\ite'$ which has
  been replaced by $\ite$ in the result of $\qu$ (see
  Example~\ref{ex:update}). 

\begin{figure}[htbp]
 \removelatexerror
\begin{algorithm}[H]
\Proc{\KwprocessQuery{\qu: Query}}{
  \KwindexQuery(\qu)\; 
  \KwmatchQuery(\qu)\; 
  \ForEach{$\ite \in$ \topk{\qu}{\ti}}{
     \Kwadd(\qu,\ite)\;
  }
} 
\Proc{\KwprocessItem{\ite: Item}}{
  \KwindexItem(\ite)\;
  \KwmatchItem(\ite)\;
  \ForEach{$\qu \in \pqueries{\ite}{\ti}^+$ }{
    \Kwadd(\qu,\ite)\;
  }
}

\Proc{\KwprocessEvent{\ev: Event}}{
  \KwmatchEvent(\ev)\;
  \ForEach{$\qu \in \pqueries{target(\ev)}{\ti}^+$ }{
    \Kwadd(\qu,\ite)\;
  }
}

\Proc{\KwmatchQuery{\qu: Query}}{
   compute  \topk{\qu}{\ti}$^+$\;
}

\Proc{\KwmatchItem{\ite: Item}}{
  compute \pqueries{\ite}{\ti}$^+$\;
}

\Proc{\KwmatchEvent{\ev: Event}}{
  compute $\pqueries{target(\ev)}{\ti}^+$; \tcp{see  Section~\ref{sec:dyn_eventHandler}}
}
\Proc{\Kwadd{\qu: Query, \ite: item}}{
  remove top-$k$ item (if it exists) from the result of $\qu$\;
  add item \ite{} to the result of query \qu \;
}

\caption{Real-time top-k query evaluation}
\label{alg:processing}
\end{algorithm}
\end{figure}

\begin{theorem}
  \label{the:processingcorrect}
  Algorithm~\ref{alg:processing} guarantees that the results of all queries are correct.
\end{theorem}
\sloppy 
\withoutproofs{The proof is omitted due to lack of space, and
  appears in the extended version~\cite{edbt16:extended}.}
\withproofs{
  \begin{proof}[sketch]
     We assume that function $\KwmatchQuery$,
    $\KwmatchItem$ and $\KwmatchEvent$ correctly compute the update
    sets $\topk{\qu}{\ti}^+$ and $\pqueries{\ite}{\ti}^+$,
    respectively, as defined in
    Section~\ref{sec:dyn_problemStatement}. 
    The correctness of the item matching
    function directly implies that all items are coherently
    \emph{added} to the query results by function $\Kwadd$. We then
    have to show that an item can only be removed from the query
    result by being replaced by an item with a higher score (function
    $\Kwadd$). This can easily be shown by the fact that we allow only
    positive event scores ($\globalscorefct$ is monotonically
    increasing).  Observe that with negative event scores, we would
    have to define a new function $\Kwdel$ which allows to replace
    some item after one or several negative scores and replace them by
    some other candidate item.
\end{proof}
}




%% file: eventHandler.tex
\section{Event Handler algorithms}
\label{sec:dyn_eventHandler}

In this section we will present two algorithms for solving the
(positive and negative) event matching problems described in the
previous section.


The first algorithm is called \emph{All Refresh} (\NAIVE). The basic
idea is to increase for each new positive event \ev{} the
query-independent score of its target item $target(\ev)$ and to
retrieve the queries whose result have to be updated,
by re-evaluating the item in the \IH.

\begin{figure}[htbp]
 \removelatexerror
\begin{algorithm}[H]
\Fn{\KwmatchEventSimple{$\ev: Event$}}{
  \If{\eventscore{\ev} == 0}{ 
    return $\emptyset$\;
  }
  $\ite:=\target(\ev)$\;
  $dynscore(\ite):=dynscore(\ite)+score(\ev)$\;
  $\KwupdateItem(\ite)$\;
  \KwmatchItem(\ite)\;
  \For{\qu $\in$ \pqueries{\ite}{\ti}$^+$}{
    \Kwadd(\qu,\ite)\;
  }
}
\caption{The $\NAIVE$ algorithm}
\label{alg:naive-algorithm-neg}
\end{algorithm}
\end{figure}

\begin{theorem}
  \label{the:naivecorrect}
  Algorithm \NAIVE{} is correct.
\end{theorem}
\sloppy
\withoutproofs{The proof is omitted due to lack of space, and appears in the extended version~\cite{edbt16:extended}.}
\withproofs{
\begin{proof}[sketch]
  We assume that function $\KwmatchQuery$, $\KwmatchItem$ and $\Kwadd$
  are correct and condition $AlgCorr(\ite)$ holds for target item
  $\ite$ before the event.  We show that for a given item \ite\ there
  exists no query $\qu \in \cqueries{\ite}{\ti}$ where
  $\globalscore{\qu}{\ite}{} > \qmin{\qu}$ after the execution of
  $\KwprocessEvent(\ev)$ processing an event \ev\ with target item
  \ite. $\KwprocessEvent(\ev)$ calls $\KwmatchItem$ (through function
  $\KwmatchEventSimple$) and then \Kwadd(\ite) after the update of the
  dynamic score.  By definition, the item matching function
  $\KwmatchItem$ returns all queries where $\globalscore{\qu}{\ite}{}
  > \qmin{\qu}$ and after processing \Kwadd(\ite), \queries(\ite)
  contains all these queries.
\end{proof}
}

The \NAIVE{} algorithm achieves an acceptable performance when events
are quite rare, compared to the item's arrival rate, but becomes
inefficient on a real-time web system suich as Twitter with feedback
events arriving from millions of users. In a highly dynamic setting we
would expect that a single event on an item, e.g. a single retweet or
favorite, should have, on average, a small impact on the set of queries
that would receive it. However, by re-evaluating the item through the
\IH, we would have to re-compute a potentially long list of candidate
queries that have already received \ite.




We propose the \emph{Real Real-Time Search} (\RRTS) algorithm, which
is based on the idea of using the \IH\ for computing \emph{for each
  item} a list of query candidates that could potentially be updated
by any future event. These lists are then processed by the \EH\ for
retrieving all queries whose result has to be updated with the item
targeted by the incoming events. Compared to \NAIVE{}, \RRTS\ avoids
examining the same list of candidate queries for an item \ite{} each
time a new event arrives targeting \ite.  The efficiency of this
algorithm relies on the following observations:
\begin{itemize}
\item each item \ite\ has a partial set of query candidates \qu\ 
  which might be updated by future events;
\item each event increases the score of a single item and triggers a
  limited number of updates
\item there exists a maximal dynamic aggregated score \thiMax{} for
  each item which depends on the scoring function and the number of
  events it will receive.
\end{itemize}


The total scoring function (Equation~\ref{eq:totalscore}) linearly
increases the item score for each new event. This leads to a
straightforward way of choosing the candidates by setting a threshold
$\Thi>0$ for each item \ite\ which controls the number of query
candidates and which is bound by the maximal dynamic score \ite\ can
receive by all future events.

\mysubsection{Candidates and threshold condition} 
Given a positive threshold value \Th, the candidates $\cands{\ite}{\Th}
\subseteq \queries$ of an item \ite\ is the set of queries:
$\cands{\ite}{\Th} = \{\qu \in \queries | \qmin{\qu} \in
(S(\qu, \ite), S(\qu, \ite) + \gamma\cdot\Th]\}$.
We say that the \emph{threshold condition} \thicondition(\ev,\Th) holds for
some event \ev{} iff the dynamic score of its target item does not
exceed threshold \Th: $ \thicondition(\ev,\Th) =
(\dynscore(\target(\ev) + \evscore(\ev)) < \Th$.
It is easy to see that when \thicondition(\ev,\Th) is true, then all
queries updated by \ev{} are in the candidate set
$\cands{\ite}{\Th}$. Otherwise, the candidate set is no longer
valid and needs to be recalculated through the \IH\ by increasing
threshold \Th.


\SetKwFunction{KwinitCandidates}{EH.refreshCandidates}
\SetKwFunction{KwinitThreshold}{EH.initThreshold}
\begin{figure}[!t]
 \removelatexerror
\begin{algorithm}[H]
\Proc{\KwprocessItem{$\ite$: Item}}{ \label{rrts:matchitem1}
  $\KwindexItem(\ite$)\; 
  $\KwmatchItem(\ite)$\;
  \For{\qu $\in$ \pqueries{\ite}{\ti}$^+$}{
    \Kwadd(\qu,\ite)\;
  }
  $\Thi:=\KwinitThreshold$\;\label{rrts:initthresh}
  $\refresh{\ite}:=0$\;\label{rrts:initrefr}
}

\Fn{\KwmatchEvent{$\ev$: Event}{: $set(Query)$}}{
  \If{\eventscore{\ev} == 0}{ \label{r2ts:sc0}
    return $\emptyset$\;
  }
  $\ite:=\target(\ev)$\; \label{r2ts:up0}
  $\dynscore(\ite):=\dynscore(\ite)+\eventscore{\ev}$\;\label{li:incdyn}
  $\KwupdateItem(\ite$)\; \label{r2ts:up1}
  \If{$\aggeventscore{\ite}{\ti} > \refresh{\ite}\cdot\Thi$}{ \label{r2ts:checkcand}\label{rrts:matchitem2}
    $\KwinitCandidates(\ite$)\; \label{r2ts:initcands1}
  } 
  \For{$\qu\ \in \cands{\ite}{\refresh{\ite}\cdot\Thi}$}{\label{beginmatch}  \label{rrts:eventmatch1}  
    \For{$\ite'\ \in \topk{\qu}{\ti}$}{
      \If{\globalscore{\qu}{\ite}{}$>$\globalscore{\qu}{\ite'}}{\label{eq:thicond}
        \Kwadd(\qu,\ite)\;\label{endmatch}  
      }
    } 
  }
}
\Proc{\KwinitCandidates{$\ite$: Item}}{ \label{rrts:refreshcands}
  $\iteThi:=copy(\ite)$\; \label{alg:placeholder}
  \If{$\Thi>0$}{
    $\refresh{\ite}:=\refresh{\ite}+\lfloor{}\aggeventscore{\ite}{\ti}/\Thi\rfloor{} + 1$\; \label{beginrefresh}  
    $\aggeventscore{\iteThi}{\ti} := \refresh{\ite}\cdot\Thi$\; \label{refresh1}
  } 
  \KwmatchItem{\iteThi}\;
  \cqueries{\ite}{\ti}:=\pqueries{\iteThi}{\ti}$^+$ - \pqueries{\ite}{\ti}\;    \label{rrts:candsins1}    \label{alg:cand1}\label{endrefresh}  
}
\caption{The \RRTS algorithm}
\label{alg:rrts-algorithm}
\end{algorithm}
\end{figure}

The threshold-based approach of \RRTS\ is detailed in
Algorithm~\ref{alg:rrts-algorithm}.  Each new item \ite\ is assigned
with a constant positive threshold \Thi\
(line~\ref{rrts:initthresh}) and a refresh counter $\refresh{\ite}$
(line~\ref{rrts:initrefr}).  $\Th=\refresh{\ite}\cdot \Thi$ represents the aggregate threshold
value after $\refresh{\ite}$ refresh steps.
When a new event arrives, function \KwmatchEvent first increases the
dynamic score (line~\ref{li:incdyn}) and updates the item index.
If the threshold condition does not hold, the candidate list is updated  by calling procedure \KwinitCandidates. The
instructions from line~\ref{beginrefresh} to line~\ref{endrefresh}
increment the refresh counter and compute the query candidates by
calling the item handler with the new aggregated threshold (item
$\iteThi$ is a placeholder copy of item $\ite$
 with a virtual aggregated event score
corresponding to the threshold of \ite\ before the next refresh). Observe that for $\Thi=0$, $\cqueries{\ite}{\ti}$ contains \emph{exactly} all queries that have to be updated by the new event.
At
line~\ref{beginmatch}  the threshold condition holds for event $\ev$ and
\KwmatchEvent copies all candidate queries \qu\ where the minimum
score of \qu\ is strictly smaller than the global score of \ite\ to
the update set \toupdate{\ev} (lines~\ref{beginmatch} to
\ref{endmatch}).


\begin{theorem}
\label{the:rrtscorrect}
Algorithm \RRTS\ is correct for positive event scores.
\end{theorem}

\withoutproofs{The proof is omitted due to lack of space, and appears in the extended version~\cite{edbt16:extended}.}
\withproofs{
\begin{proof}[sketch]
\ber{CHECK / SIMPLIFY PROOF}
  We have to show that both functions, $\KwprocessItem$ and
  $\KwmatchEvent$, guarantee that for all items $\ite$ and all
  queries $\qu\in \queries$, if $\globalscore{\qu}{\ite}{} >
  \qmin{\qu}$, then $\qu \in \pqueries{\ite}{\ti}$ (condition $AlgCorr$). 
  It is easy to see that the correctness condition holds for some new
  item $\ite$ after the execution of function $\KwprocessItem$ (proof
  of Theorem~\ref{the:processingcorrect}).
  
  We now show that $AlgCorr$ also holds after the execution of
  function $\KwmatchEvent$.  If $\eventscore{\ev}=0$ the result does
  not change (line~\ref{r2ts:sc0}).  Otherwise, lines~\ref{r2ts:up0}
  to~\ref{r2ts:up1} update the event score of target item
  $\target(\ev)$.
Line~\ref{r2ts:checkcand} checks if the
candidate set $\cands{\ite}{\refresh{\ite}\cdot\Thi}$ of item \ite{}
has to be updated.
If this is the case, lines~\ref{beginrefresh} to~\ref{endrefresh}
increment \refresh{\ite} by one and recompute the candidate set
$\cands{\ite}{\refresh{\ite}\cdot\Thi}$.  We know at
line~\ref{beginmatch} that
  the candidate set $\cands{\ite}{\refresh{\ite}\cdot\Thi}$ of \ite\ contains \emph{all} queries
  $\qu \in \queries-\pqueries{\ite}{\ti}$ where
  \begin{eqnarray}
    \qmin{\qu}\leq \globalscore{\qu}{\ite}{} = \alpha \cdot \queryscore{\ite}{\qu} + (1-\alpha)\cdot\refresh{\ite}\cdot\Thi
    \label{eq:cond2}
  \end{eqnarray}


  Then we can show that $\cands{\ite}{\refresh{\ite}\cdot\Thi}$
  contains all queries that have to be updated by event $\ev$ (and
  other queries which are not updated). By definition and
  equation~\ref{eq:cond2}, for all queries $\qu'$ to be updated by
  event $\ev$ the following holds
  \begin{eqnarray*}
    \label{eq:5}
    \qmin{\qu'} < \globalscore{\qu'}{\ite}{}&=&\statscore(\qu', \ite) + \gamma\cdot\aggeventscore{\ite}{\ti}\\
   &\leq&\statscore(\qu', \ite) + \gamma\cdot\refresh{\ite}\cdot\Thi    \label{eq:cond10}
  \end{eqnarray*}
  
  which means that $\qu'\in \cands{\ite}{\refresh{\ite}\cdot\Thi}$ and
  if $\globalscore{\qu}{\ite}{} > \qmin{\qu}$, then $\qu \in
  \pqueries{\ite}{\ti}$ after executing lines~\ref{beginmatch}
  to~\ref{endmatch} (condition $AlgCorr$).
\end{proof}
}

The challenge arising from this algorithm is twofold. Identify optimal
threshold values \Thi\ for each item and index the obtained candidate
queries in a way that allows efficient retrieval of updates during the
event matching operation (see Section~\ref{sec:dyn_candIndexing}).

\mysubsection{Cost analysis} The choice of threshold \Thi\ controls
the number of candidate query refresh operations and has an important
impact on the overall performance of the system.  


  From Algorithm~\ref{alg:rrts-algorithm}
we can easily understand that higher values of \Thi\ minimize the
number of costly candidate refresh operations (by calling the \IH) but
also might generate a large number of false positive query candidates
which will never be updated by an event targeting $\ite$. On the other
hand, low values of \Thi\ lead to more frequent costly candidate
re-evaluations on the \IH.
\ber{Do we really see this in Section 5?: As we will see in
Section~\ref{sec:dyn_experiments} the total execution time for item
and event evaluation on Twitter's stream can differ to up to an order
of magnitude, based on the choice of \Thi.}

\longversion{
Ideally, a perfect estimation on the item's maximal dynamic score item
would enable to match an item only once in the \IH\ and then, all
incoming events could be matched in the \EH{} to refine the list of
candidate queries. However, according to our empirical evaluation in
Section~\ref{sec:dyn_experiments} this ideal solution might still not
be optimal due to the higher average matching and maintenance cost for
large candidate lists in the \EH.
}
%
\newcommand{\CEIL}[1]{\ensuremath{\left \lceil{#1}\right \rceil}} %
\newcommand{\numEvalIH}{\ensuremath{\CEIL{\thiMax / \Thi}}} %
\newcommand{\NumEvalIH}{\ensuremath{\CEIL{\frac{\thiMax}{\Thi}}}}
%


In the following we will use the  notation in Table~\ref{tab:costpars}.
\begin{table}[htbp]
  \centering
  \begin{tabular}[htbp]{|c|p{5.5cm}|}
\hline
  $\ite$&target item\\
\hline
  $\Thi$&estimated threshold value for $\ite$\\
\hline
$\thiMax$& estimated maximal aggregated event score for $\ite$\\
\hline
  $\evMax$&estimated total number of events for $\ite$\\
\hline
  $\rr$&estimated number of candidate refresh operations  for $\ite$\\
\hline
$\cost{\IHshort}(\ite, \Thi)$&aggregated \emph{item matching  cost}\\
\hline
$\cost{\EHshort}(\ite, \Thi)$&aggregated \emph{event matching  cost}\\
\hline
$\cands{\ite}{}$&average size of the candidate list of item $\ite$.\\
\hline 
$\cost{T}$&cost of checking if item $\ite$ updates candidate $\qu$\\
\hline
$\cost{M}(\ite)$& average item matching cost  for $\ite$\\
\hline
$\cost{C}(\ite, \Thi)$& average candidate list construction cost for $\ite$ and $\Thi$\\
\hline
$cost(\ite, \Thi)$&aggregated \emph{total matching cost}\\
\hline
\end{tabular}

\caption{Cost function parameters}
\label{tab:costpars}
\end{table}
%
%
Under the assumption that each event targets only one item and that
items are processed independently, we are interested in finding an
optimal value for the threshold value $\Thi$ that minimizes the
aggregated total cost $cost(\ite, \Thi)$ for each item $\ite$.  This
local optimization leads to a globally optimized evaluation cost.
%

By definition the total aggregated matching cost is defined by the sum
of the aggregated item and the aggregated event matching costs:
 $$cost(\ite, \Thi)=\cost{\IHshort}(\ite, \Thi)+\cost{\EHshort}(\ite, \Thi)$$

 If we suppose that all events have the same event score
 \eventscore{\ev}, then $\rr =\numEvalIH$.
\longversion{Observe also that for a positive threshold $\Thi$ which is smaller
 than the \eventscore{\ev}, $0< \Thi<\eventscore{\ev}$, the number of
 events is smaller than the number of candidate refresh operations:
 $\evMax < \rr < \numEvalIH$.}


%
%

We can estimate the aggregated item and event matching costs
$\cost{\IHshort}(\ite, \Thi)$ and $\cost{\EHshort}(\ite, \Thi)$ as
follows (Algorithm~\ref{alg:rrts-algorithm}).  First, each candidate
list is refreshed $\rr$ times where each refresh consists in matching
the item and constructing the candidate list (function \KwinitCandidates):
\begin{equation}
  \label{eq:costIH}
  \cost{\IHshort}(\ite, \Thi) = \rr \cdot (\cost{M}(\ite) + \cost{C}(\ite, \Thi))
\end{equation}
Second, each event is matched in the \EH\ $\evMax$ times. Supposing
that there is \emph{no early stopping condition} when checking the
candidates, the cost of every event match operation depends on the
average size of the candidate list
$\cands{\ite}{\refresh{\ite}\cdot\Thi}$ and on the (constant) cost
\cost{T} of checking if item $\ite$ updates a given query candidate
$\qu$:
 \begin{equation}
  \label{eq:costEH}
  \cost{\EHshort}(\ite, \Thi) = \evMax \cdot \cands{\ite}{\refresh{\ite}\cdot\Thi}\cdot \cost{T}
  \end{equation}

\longversion{
When \Thi\ = 0, no
candidates are maintained and all events are matched in the \IH. For
$\Thi=0$, $\rr=\evMax$, $\cost{C}(\ite, \Thi)=0$ and
$\cands{\ite}{\refresh{\ite}\cdot\Thi}=0$, we obtain
$\cost{\IHshort}(\ite, \Thi) = \evMax \cdot \cost{M}(\ite)$  and 
$\cost{\EHshort}(\ite, \Thi) = 0$.
%
When $\Thi = \thiMax$, all event evaluations are performed on the \EH\ 
($\rr=1$):
$\cost{\IHshort}(\ite, \thiMax) = \cost{M}(\ite) +  \cost{C}(\ite, \Thi)$ and 
$\cost{\EHshort}(\ite, \thiMax) = \evMax \cdot
  \cands{\ite}{\refresh{\ite}\cdot\Thi} \cdot \cost{T}$.
}

\mysubsection{Finding optimal \bfseries{\Thi}}
%
%
%
%
%
The value of $\cands{\ite}{\refresh{\ite}\cdot\Thi}$ depends on the
distribution of the query minimum scores. Our cost model relies on the
assumption that the average size of all candidate lists produced by
each refresh step linearly depends on threshold value $\Thi$:
$\cands{\ite}{\refresh{\ite}\cdot\Thi} = a\cdot \Thi$, where $a$ is a
positive constant\footnote{A more precise cost model estimating the
  optimal threshold \emph{for each individual refresh} would need a
  query/item/event distribution model which is difficult to obtain for
  a real-world workload.}. Similarly, we assume that the candidate creation
\cost{C}(\ite, \Thi) depends linearly on \Thi: $\cost{C}(\ite,
\Thi)=b\cdot\Thi$. This cost abstractions simplify the computation of
the optimal \Thi and they have been also experimentally validated.

Given $\rr =\thiMax/\Thi$ (we consider here the real value instead of
the integer floor value),  Equations (\ref{eq:costIH}) and (\ref{eq:costEH}) the
total $cost(\ite, \Thi)$ can be written as:

\begin{eqnarray*}
  cost(\ite, \Thi) &=& \frac{1}{\Thi}\cdot\overbrace{\thiMax \cdot \cost{M}(i)}^{c1} 
  \\ && +   \Thi \cdot \underbrace{\evMax  \cdot a \cdot \cost{T}}_{c_2}  
  + \underbrace{\thiMax \cdot b}_{c_3}
  \end{eqnarray*}
  
We minimize the $cost$ function using the first derivative:

\begin{equation*}
\frac{d(cost(\ite, \Thi))}{d\Thi}= -\frac{c_1}{(\Thi)^2}  + c_2
\end{equation*}
Function $cost$ is monotonically
decreasing in the interval $(0, \sqrt{c_1 / c_2})$ and increasing in
$(\sqrt{c_1 / c_2}, \thiMax]$, thus making $\Thi = \sqrt{c_1 / c_2}$
the optimal value:
  
$$\Thi^{opt}=\sqrt{\frac{1}{a} \cdot \frac{\thiMax}{\evMax } \cdot  \frac{\cost{M}(i)}{\cost{T}}}$$

\ber{I did a more rich conclusion} This equation essentially shows
that $\Thi^{opt}$ depends on the query distribution in the \IH{} index
(first factor) and increases with the ratio $\cost{M}(i)/\cost{T}$
between the item matching cost and the event test cost (third factor).




%% file: candidateIndexing.tex
\section{Candidate indexing}
\label{sec:dyn_candIndexing}

The \EH\ matches incoming events against a precomputed list of query
candidates which has to be regularly refreshed. In the above cost
model, we do not take account of the cost of maintaining the candidate
list. This cost clearly depends on the size of the list, as well as,
on the choice of the data structures for indexing candidates in the
\EH. In this section we introduce three indexing schemes aiming to
optimize the cost of storing and retrieving candidates. We are
particularly interested in finding efficient early stopping conditions
during event matching in order to avoid visiting all candidates.

The main task of the \EH\ is to retrieve result updates triggered by
any arriving event. Since each event \ev\ increases the score of a
single item \target(\ev), these updates will only concern a subset of
this item's candidate queries. Hence, the building block of all
the indexes we propose is a dictionary from each item to its set of
candidates. The proposed indexes aim at organizing the \emph{posting
  lists}, i.e. the candidates sets in a way that decreases the number
of false positives encountered during event matching and to efficiently support insertions and deletions. 
\longversion{\begin{figure}  
   \centering 
   \begin{subfigure}[h]{\linewidth}
     \centering
     \includegraphics[width=.8\linewidth]{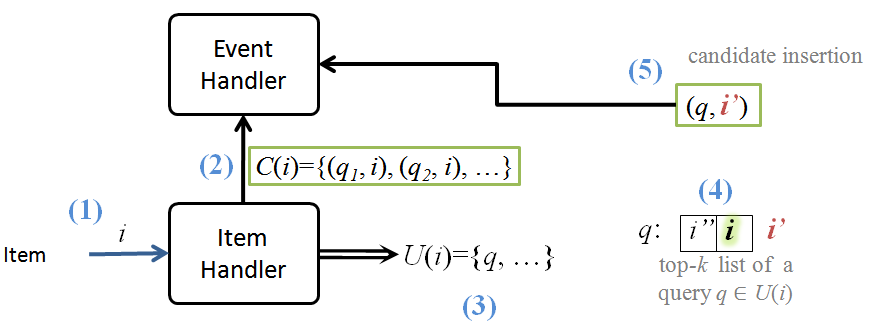}
     \caption{An item \ite\ (1) is evaluated by the \IH\ and a set of candidates is computed (2). This set is inserted into the \EH. The updates \toupdate{\ite} are also retrieved (3). The insertion of \ite\ in a query \qu\ triggers the deletion of another item \ite' from its top-$k$ list (4). If \ite' fulfills the \Thi-condition, then \qu\ is inserted as its candidate (5).}
   \end{subfigure}
   
   \centering 
   \begin{subfigure}[h]{\linewidth}
     \centering
     \includegraphics[width=.8\linewidth]{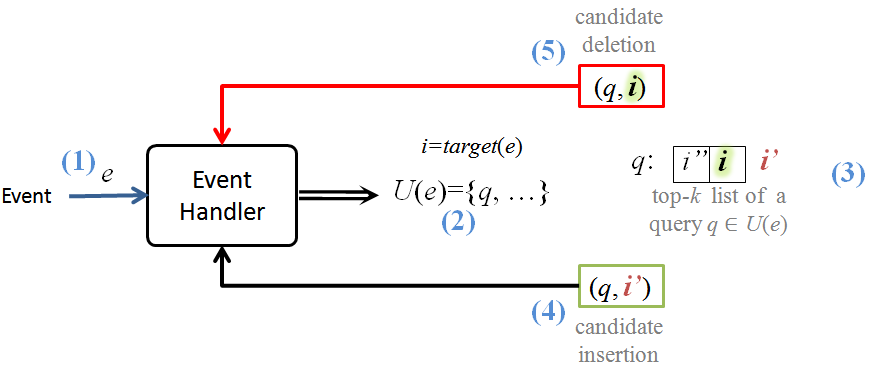}
     \caption{An event \ev\ (1) is evaluated in the \EH\ and the set of updates is computed (2). As before, any deleted item \ite' (3) can potentially be inserted in the Event Handler (4). At the same time, since \ite\ is inserted in the top-$k$ list of \qu, \qu\ is no longer its candidate and has to be removed (5)}
   \end{subfigure}
\captionsetup{justification=centering}
\caption[Updating query candidates]{Insertion and deletion of query candidates}
\label{fig:candUpdates}
\end{figure}
\emph{Candidate insertions} occur during candidate list refresh
operations (function \KwinitCandidates) or  query result updates.  In the latter
case, when an item $\kite$ is deleted from the top-$k$ list of a query
\qu, \qu\ is added to the candidate list of $\kite$ in the \EH{} by
procedure $\Kwadd$.  \emph{Candidate deletions} may occur during event
matching in two situations
. The first situation arises when a query update is detected while
iterating through the candidate set after the arrival of a new event
(lines~\ref{rrts:eventmatch1} to \ref{endmatch} in
Algorithm~\ref{alg:rrts-algorithm}). Then this query should be deleted
from the candidate list.  A second case is when a candidate query \qu\
achieves a minimum score higher than the \Thi\ threshold of some item
\ite. It is obvious that in this case it is safe to remove \qu\ from
the candidates set of $\ite$ (this optimization is not part of
Algorithm~\ref{alg:rrts-algorithm}).
}
Although deletion in both cases in not necessary for the correctness
of the algorithm, maintaining these queries as candidates would lead
to an unnecessary number of false positives, deteriorating the overall
event-matching performance for following event evaluations.
\begin{figure}[htbp]   
   \begin{subfigure}[h]{.28\linewidth}
     \includegraphics[width=\linewidth]{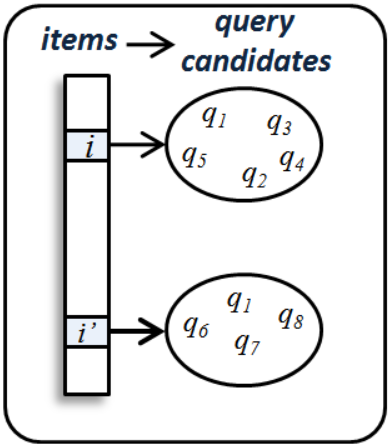}
     \caption{}
   \end{subfigure}
   ~
   \begin{subfigure}[h]{.28\linewidth}
     \includegraphics[width=\linewidth]{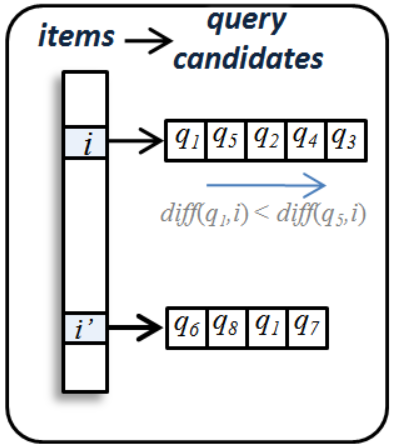}
     \caption{}
   \end{subfigure}
   ~
   \begin{subfigure}[h]{.29\linewidth}
     \includegraphics[width=\linewidth]{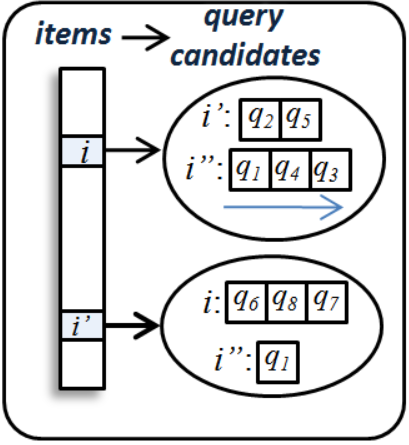}
     \caption{}
   \end{subfigure}
\captionsetup{justification=centering}
\caption[Event Handler indexes]{(a)Simple (b)Ordered (c)Partitioning indexes}
\label{fig:dyn_dataStructures}
\end{figure}
%
A straightforward implementation of the \EH\ posting lists is to
maintain an unordered set of candidates as illustrated in
Figure~\ref{fig:dyn_dataStructures}-(a). Assuming a dynamic array
implementation of this index, insertions and deletions of candidates
can be performed in (amortized) constant time.  On the one hand, if
$\Thi$ is large enough, this solution has the advantage of avoiding
candidate list updates for a potentially big number of arriving events
(as long as the \Thi-condition holds). On the other hand, it is not
possible to apply any early stopping condition, and all candidate
queries need to be checked for
updates.  

Following the \ber{I'm not sure if we should reference Fagin here:
  Fagin's Threshold Algorithm~\cite{Fag03}}  ``sort and prune''
principle, we try to define an efficient ordering and appropriate data
structures for matching query candidates that will allow us to define
an early stopping condition
(Figure~\ref{fig:dyn_dataStructures}-(b)). 
As defined in Section \ref{sec:dyn_eventHandler}, a query \qu\ is a
candidate of an item \ite\ if $\textit{diff}(\qu,\ite)=\qmin{\qu} -
\score(\qu, \ite) \in (0, \Thi]$ and $\textit{diff}(\qu,\ite)$
represents the lower dynamic score bound \ite\ must receive before
updating \qu ($\qmin{\qu}$ is monotinically increasing). So, if
candidates are ordered by value $\textit{diff}(\qu,\ite)$, on event
arrival, it is sufficient to visit only candidates with
\emph{negative} $\textit{diff}(\qu,\ite)$. Notice that this set of
visited candidates guarantees not only zero false negatives but also
zero false positives. Nevertheless, maintaining this order in a
dynamic environment is a non-trivial problem: from the moment a query
becomes a candidate of an item \ite{} until to a new event with target
\ite, the minimal scores of other candidates might have changed
(increased) due to result updates or, more frequently, due to events
received by their $k$-th item. In the following, we will discuss the
implications of maintaining the accurate order and present two
lazy re-orderinf approaches reducing update costs.

\mysubsection{Exhaustive ordering Index} 
This solution accurately maintains the order of query candidates by
the score difference $\textit{diff}(\qu,\ite)$. To achieve this, it
executes all necessary re-orderings on each \qmin{\qu} change,
independently to whether it is due to an event or an item and in all
postings of items that \qu\ is a candidate.
\longversion{ To better understand the high maintenance cost of this
  index, consider that an event \ev\ arrives and is matched through
  the \EH. Since the ordering is correct, we can immediately detect
  the correct list of updates. However, to maintain the invariant of
  the index the algorithm must find all queries \qu\ that currently contain
  this item \ite\ as their $k$-th element (determining the
  \qmin{\qu}). For each such query \qu, we need to find all items
  \ite' where \qu\ is a candidate and then re-order them. The same
  procedure needs to be followed also when a query \qu\ is updated by
  an item \ite.  }
We will observe in our experiments that in a real-world workload
the performance gains from avoiding false positives are outperformed
by the time wasted to maintain the order. 

The following two solutions follow a lazy re-ordering approach, in
order to achieve a reasonable trade-off between the cost of
eliminating false positives and the cost of re-orderings
candidates. 
\longversion{In both solutions, we consider that for each candidate, we
store an additional value representing $\textit{diff}(\qu,\ite)$, at
the time of the candidates indexing.}

\mysubsection{Static-order Index} 
This solution sorts all candidates only at the creation of the candidate list.  
New queries are inserted to the beginning of the candidate lists,
independently of their $\textit{diff}(\qu,\ite)$ values. All minimal
score \qmin{\qu} changes are also ignored. On event arrival, the
matching algorithm vists all candidate queries for which the
\emph{stored} (and not actual) difference score is lower than the
item's dynamic score:
$\textit{diff}(\qu,\ite)<\aggeventscore{\ite}{\ti}$. It is easy to
show that this condition is safe but also converges with time to the
previous simple \EH\ solution.

\mysubsection{Lazy re-ordering Index} The two previous approaches
either have a very high maintenance cost due to frequent re-ordering
operations (\emph{Exhaustive}), or an inefficient event matching due
to lack of maintenance of this order (\emph{Static-order}). The Lazy
approach lies in-between by following a lazy ``on false-positive''
update approach. Given an item \ite, any order changes caused by
minimal score updates are ignored until the next arrival of a new
event with target $\ite$.  The heuristics behind this approach is that
many re-orderings of the \emph{Exhaustive} approach never are
exploited because candidate queries are moved several times before
actually being visited on an event arrival. Therefore, the \emph{Lazy}
approach attempts to minimize the number of re-orderings at the expense
of introducing false positives on event matching. Given that in the
general case the cost of re-ordering an element in a sorted list has a
logarithmic complexity, while the cost of eliminating a false positive
is constant, the \emph{Lazy} approach is expected to outperform the
\emph{Exhaustive} one.

\mysubsection{Item Partitioning Event Handler} Ordered indexes reach
their limits for highly varying $\textit{diff}(\qu,\ite)$ values.
\longversion{ When studying \emph{Exhaustive}, we observe that
  frequent re-orderings in the posting lists can lead to a rather
  inefficient performance. Even in the more efficient,
  \emph{Exhaustive}, the logarithmic complexity of re-orderings can
  result to a poor performance in cases of a high number of false
  positives.  } The \emph{Item Partitioning} approach relies on the
following observation: if two queries $\qu_1$ and $\qu_2$ are both
candidates of an item \ite\ and both have the same item \ite' as their
$k$-th result item, both difference score values
$\textit{diff}(\qu_1,\ite)$ and $\textit{diff}(\qu_2,\ite)$ are
synchronized and the relative order of both queries in the posting
list of $\ite$ will remain the same. For this reason, we organize the
query candidates of each item, with respect to their $k$-th
element. The order assigned in each such group during insertion
remains constant for as long as their $k$-th result remains the same,
i.e. as long as they do not receive any updates.  For example, in
Figure~\ref{fig:dyn_dataStructures}, item \ite\ has 5 candidate
queries, two of which ($\qu_2$ and $\qu_5$) have \ite' as their $k$-th
element. These queries are grouped together
(Figure~\ref{fig:dyn_dataStructures}-(c)) in a sorted list, (ordered
according to $\textit{diff}$). For as long as \ite' remains the $k$-th
element of both these candidate queries, the ordering of the list is
static. On event arrival, the matching operation checks each one of
these groups if the corresponding item, until reaching true negative
candidate. In case of an update, however, of a query \qu\ by an item,
\qu\ has to be re-indexed in the group of its new $k$-th element in
all items where it is a candidate. Despite the additional cost on the
query update operation, as we will see later in
Section~\ref{sec:dyn_experiments}, the minimization of both
re-orderings (only on the case of updates) and false positives leads
to an overall better performance than the previous approaches.


%% file: experiments.tex
\section{Experiments}
\label{sec:dyn_experiments}
\newcommand{\indNaive}{\textsc{Naive}}
\newcommand{\indSimple}{\textsc{Simple}}
\newcommand{\indLazy}{\textsc{LazyOrder}}
\newcommand{\indGraph}{\textsc{ItemPart}}

In this section we experimentally evaluate the algorithms presented in
Section~\ref{sec:dyn_eventHandler} and the data structures proposed in
Section~\ref{sec:dyn_candIndexing}, using a real dataset collected
from Twitter's public streams. Through these experiments we compare
the performance of the \RRTS\ algorithm implementations over the
simple candidates (\indSimple), the lazy ordering (\indLazy) and the
item partitioned (\indGraph) index with the na\"{i}ve approach of the
\NAIVE\ algorithm (\indNaive) over a number of parameters, like the
total number of stored continuous queries, the size $k$ of query
result lists, and the weight of user feedback ($\gamma$) on the total
score function. Additionally, we are interested in assessing the
impact of the \Thi\ tuning parameter over the overall system
performance.  \nel{remove? We present the execution time required to
  evaluate the items and user events and also, the pruning gains
  achieved by each implementation over the list of candidates, as a
  percentage of the visited queries, before meeting the stopping
  condition requirements.}  Our experiments rely on a slightly
modified version of the item matching algorithm\footnote{Available as
  an open source library \texttt{continuous-top-k}:
  https://code.google.com/p/continuous-top-k/} (\IH{}) presented in
\cite{VAC12} to additionally detect candidate queries given a value of
\Thi.

\begin{table}[htbp]
\centering
\begin{tabular}{ l|r|r|r|r}
              &  \#items    & \#events   & min  & avg  \\\hline
DS1           & 10 676 097  & 13 787 349 &  1              &  1.29 \\
DS5 (default) &    201 581  &  2 013 427 &  5              &  9.99 \\
DS10          &     56 417  &  1 105 639 & 10              & 19.60
\end{tabular}
\caption{Number of items and events in each dataset}
\label{tab:dyn_datasets}
\end{table}
\begin{table}[htbp]
\centering
\begin{tabular}{ r|r|r}
  parameter   & default value & range \\\hline\hline
  \#queries   & 900 000       & [100 000, 900 000] \\\hline
  $\alpha$    & 0.3           & $(1 - \gamma) / 2$ \\
  $\beta$     & 0.3           & $(1 - \gamma) / 2$ \\
  $\gamma$    & 0.4           & $[0.05, 0.95]$ \\\hline 
  $k$         & 1             & $[1, 20]$ \\\hline
  $\Thi$ strategy  &  $\Thi^{max} / 2$    & $[0, 0.2]$ or $[0, \Thi^{max}]$ \\
\end{tabular}
\caption{Experimental parameters}
\label{tab:dyn_expParams}
\end{table}
Experiments were conducted using an Intel Core i7-3820 CPU@3.60.
Algorithms were implemented in Java 7 and executed using an upper
limit of 8GB of main memory (-Xmx8g). The configuration used only one
core for the execution of the different algorithms. We present the
\emph{average execution times of three identical runs} after an
initial warm-up execution. All queries were stored before any item or
event evaluation and their insertion time is not included in the
results. We collected a real-world dataset from the Twitter Stream API
over a period of 5 months (from March to August 2014). In our data model,
an original tweet corresponds to an item and a retweet to a feedback
signal, i.e. an event for the original item. From this set we have
filtered out non-English tweets, as well as those without any retweets
(events), leading to a dataset of more than 23 million tweets and
retweets (DS1) (Table~\ref{tab:dyn_datasets}). Two additional subsets
were created by considering only tweets (and the corresponding
retweets) with at least 5 (DS5) or 10 retweets (DS10) per item.  DS5
dataset is the default dataset for the experiments and contains 2.2
million tweets and retweets.  Queries were generated by uniformly
selecting the most frequent n-grams of 1, 2 or 3 terms from the tweet
and retweet dataset leading to an average length of 1.5 terms per
query.  In each of the following experiments we change one parameter
within a given range, while all system parameters remain constant.
Table~\ref{tab:dyn_expParams} shows the default values for each
parameter, as well as the range of each parameter used for the
corresponding experiments.

\begin{figure*}[btp]
  \centering
  \begin{subfigure}[t]{.32\linewidth}
    \includegraphics[width=\columnwidth]{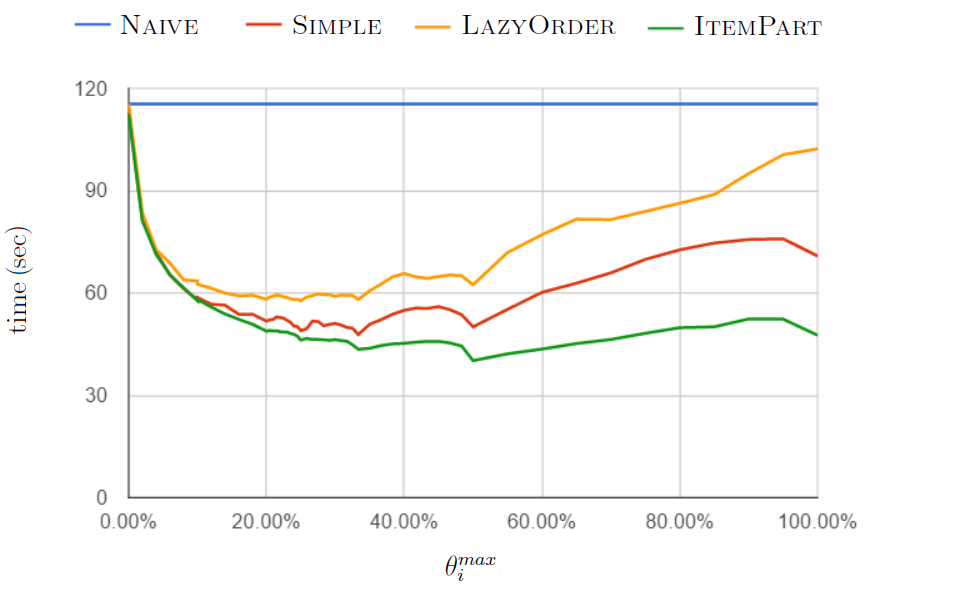}
  \caption{Time exact threshold $\Thi^{max}/Th$}
    \label{fig:dyn_onThetaMax}
  \end{subfigure}
  \begin{subfigure}[t]{.32\linewidth}
    \includegraphics[width=\linewidth]{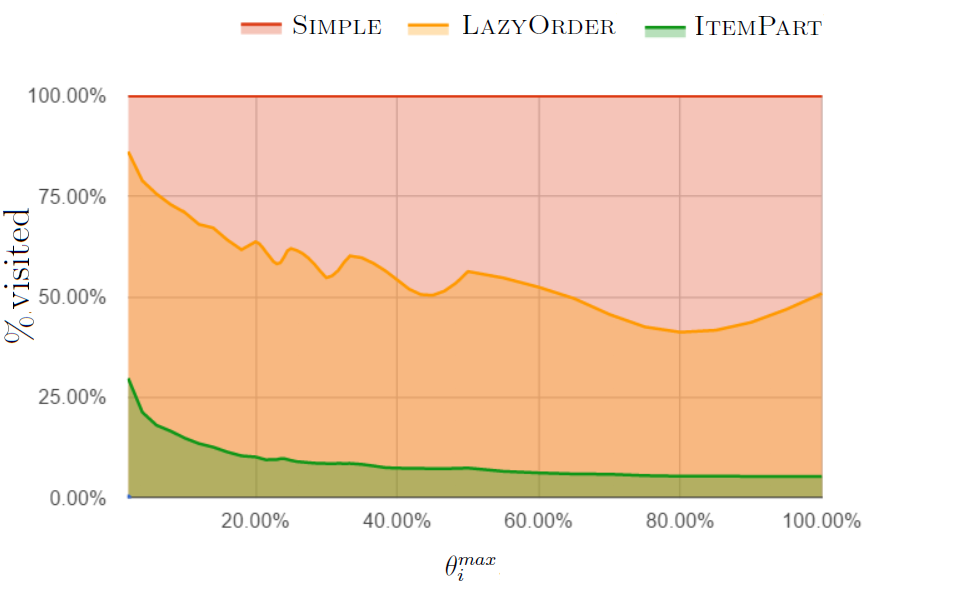}
      \caption{Selectivity exact threshold $\Thi^{max}/Th$}
    \label{fig:dyn_onThetaMax_visited}
  \end{subfigure}%
  \begin{subfigure}[t]{.32\linewidth}
    \includegraphics[width=\linewidth]{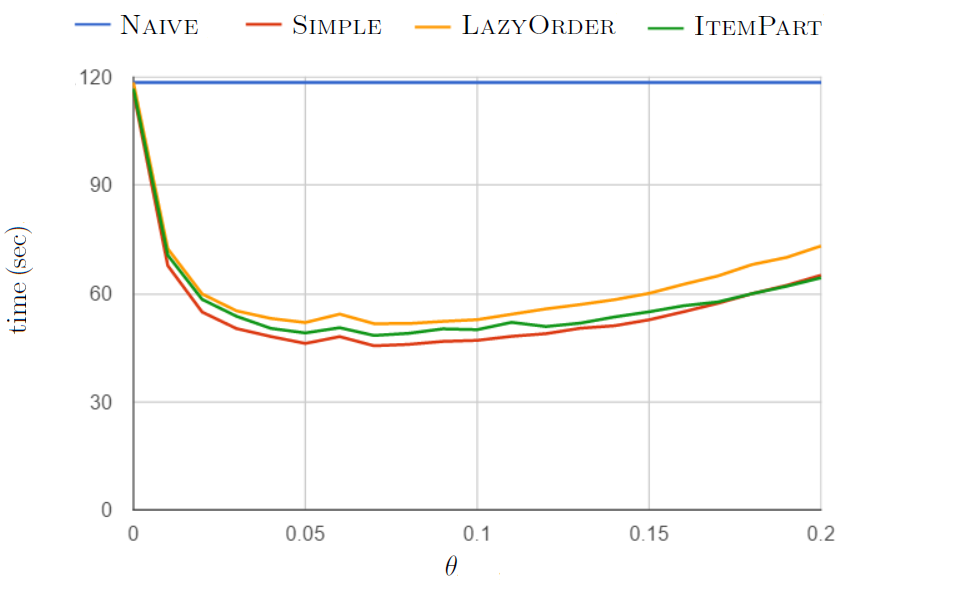}
    \caption{Time global threshold value $\Th$}
    \label{fig:dyn_onTheta}
  \end{subfigure}
  \label{fig:rw_time}
\caption{Exact and Global Threshold}
\end{figure*}
\mysubsection{Exact maximal threshold $\Thi^{max}$} In this experiment
we use as maximal threshold $\Thi^{max}$ the exact final aggregated
event score \aggeventscore{\ite}{\ti}
(Equation~\ref{eq:aggeventscore}) of each item \ite. The horizontal
axis of Figure~\ref{fig:dyn_onThetaMax} represents the percentage
(from 0 to 100\%) of the $\Thi^{max}$ value assigned as \Thi\ value to
an incoming item while the vertical axis represents the time required
to match the items and events of the dataset DS5. \indNaive\ execution
time is independent of the \Thi\ value and thus shown as a constant
line. Note that in the special case of $\Thi=0$ all indexes converge
to the \indNaive\ one.  \longversion{ : $\Thi=0$ implies that no 
  candidates will be maintained and thus, all event evaluations will
  be performed in the \IH, which is exactly the behavior of
  \indNaive.}  \indGraph\ outperforms the other three indexes, while
\indLazy\ has higher execution time than \indSimple, despite the early
stopping condition defined. The relatively good performance of
\indSimple\ is attributed to the low maintenance cost (no
re-orderings) and the use of a dynamic vector data structure, which
guarantees fast access, insertions and deletions. These factors
compensate for the lack of an early stopping condition.  \indLazy\ and
\indGraph\ use much ``heavier'' data structures and need to prune a
large portion of the candidates lists in order to outperform
\indSimple.

We can also observe a particular pattern in this plot: execution time
exhibits local minima for $\Thi$ values which are fractions of 
$\Thi^{max}$, i.e.,  100\%, 50\%, 33\%, 25\% etc. To understand this form, 
consider for example the case of $\Thi = 0.5 \cdot \Thi^{max}$. On the 
first evaluation of any given item \ite, the list of candidates with a 
difference up to \Thi\ is computed. Since \Thi\ corresponds to half the 
maximal value of the aggregated item score, the item will be re-evaluated 
through the \IH\ a second time, and there will be no need for a third 
evaluation. When a higher value is chosen, e.g., $0.6 \cdot \Thi^{max}$, 
two evaluations in the \IH\ would also be required, with the difference 
that some already known candidates would be retrieved again. Hence, an 
additional cost incurs to a) compute the redundant candidates and b) to
filter out the probably higher number of false positives on event
matching in the \EH. 

Figure \ref{fig:dyn_onThetaMax_visited} shows the average percentage
of lists visited, until the stopping condition becomes true. The size
of the candidate lists is the same for all algorithms for the same
\Thi{} threshold. \indSimple\ always visits 100\% of the lists due to
the lack of a stopping condition. We can observe that \indGraph\ has a
20\% smaller execution time (Figure~\ref{fig:dyn_onThetaMax}) while
visiting in average only 5\% of the candidate lists.
\begin{figure*}[tbp]
  \centering
  \begin{subfigure}[t]{0.32\linewidth}
    \includegraphics[width=\linewidth]{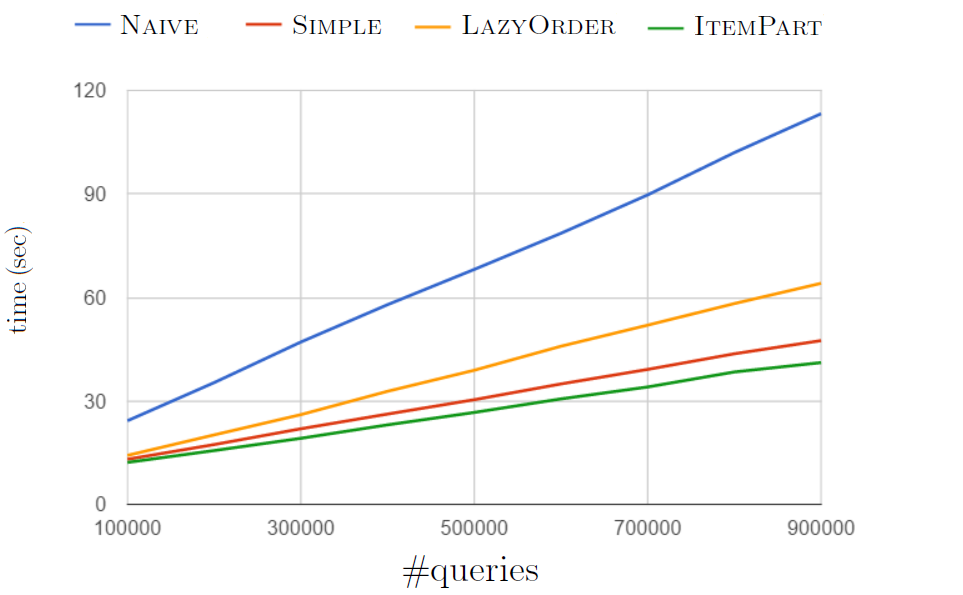}
    \caption{Number of queries}
    \label{fig:dyn_onQueries}
 \end{subfigure}
\begin{subfigure}[t]{0.32\linewidth}
    \includegraphics[width=\linewidth]{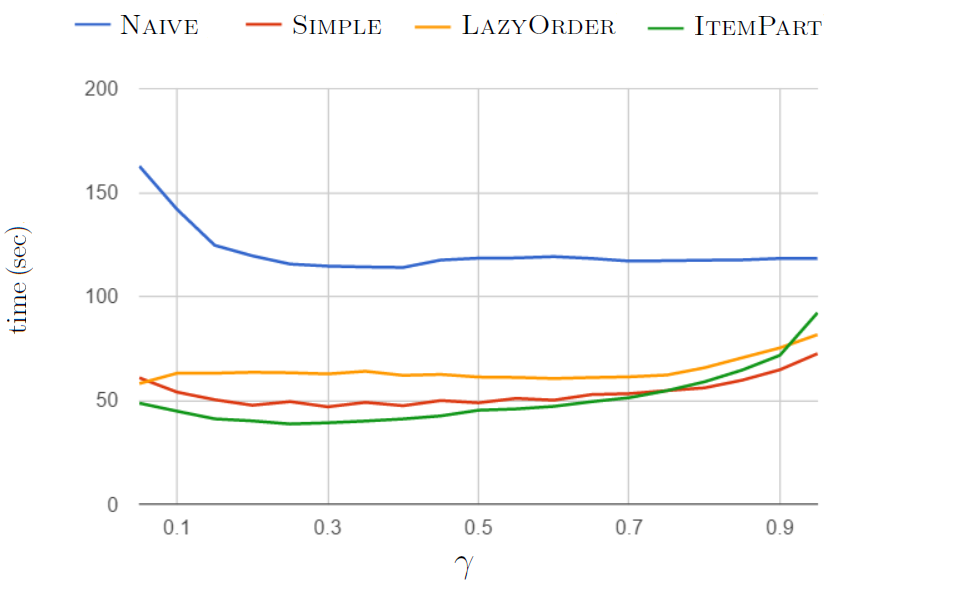}
    \caption{Dynamic score weight $\gamma$}
    \label{fig:dyn_onGamma}
\end{subfigure}
\begin{subfigure}[t]{0.32\linewidth}
    \includegraphics[width=\linewidth]{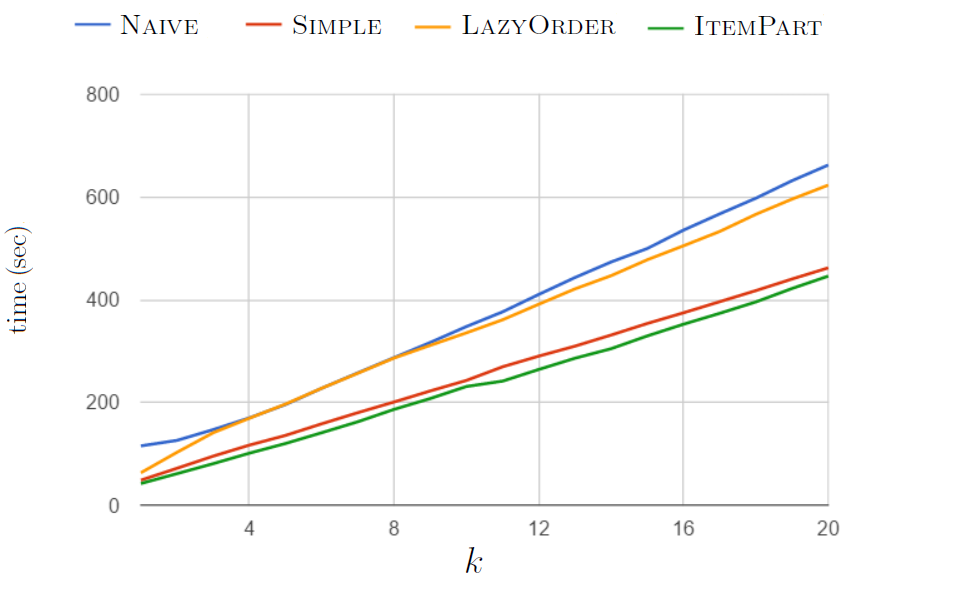}
    \caption{Result parameter $k$}
    \label{fig:dyn_onK}
\end{subfigure}
\caption{Scalability}
\end{figure*}

\mysubsection{Global threshold $\Th$} \longversion{The previous experiment relies
on a perfect estimation of the $\Thi^{max}$ value, which is a rather
unrealistic assumption in a high dynamic environment with complex
query/item/event workload distributions.} In this experiment we assign
\emph{the same \Th\ value to all items} without using any knowledge of
the maximal dynamic score per item. Figure~\ref{fig:dyn_onTheta}
shows, for each \emph{absolute value} \Th\ assigned to all items, the
time required to evaluate the whole dataset (DS5). We can observe that
as the value of \Thi\ increases from 0, there is a quick improvement
in the performance of the indexes \indSimple, \indLazy\ and
\indGraph\, while after exceeding the optimal \Thi\ point, it
deteriorates with a smaller slope: for very small values of \Th, the
lists become frequently obsolete (when the \Thi-condition no longer
holds) and a large number of incoming events need to be evaluated in
the \IH. This explains the first phase where execution time
decreases. As values of \Th\ become much bigger, the lists become
longer: computing the candidates becomes more costly and more false
positives are likely to appear. Unlike in the previous experiment the
three indexes exhibit similar performance with no more than 5\% of
difference in execution time.  This behavior indicates that the
stopping condition for both \indLazy\ and \indGraph\ fails to prune as
much query candidates as in the previous experiment. This is
attributed to the arbitrary assignment of the same \Th\ value to all
items. For some items this value can be large w.r.t.  $\Thi^{max}$ and
the algorithms thus spend unnecessary time for finding candidate
queries which will never be used. On the other hand, a small value of
the \Thi\ w.r.t. $\Thi^{max}$ means that there will be to many costly
evaluations of events in the \IH.

\mysubsection{Scalability and throughput} In this experiment, we are
interested in the scalability of our indexes w.r.t. the number of
continuous queries stored in the system (index creation time is
excluded from our measurements). As we can see in
Figure~\ref{fig:dyn_onQueries} all implementations scale linearly with
the number of continuous queries.  \longversion{ : as more queries
  stored, the size of posting lists, in the \IH\ and of candidate
  queries lists also increases demanding more time to iterate through
  the indexes to retrieve the updates.}  \indGraph\ scales better:
over $100,000$ queries it requires 50\% of the corresponding time for
\indNaive\, while over $900,000$ it only requires 36\%. The low slope
of \indGraph\ indicates the good performance of its stopping condition
to prune more candidate queries over increasing list sizes.
\longversion{ This observation can also be made by looking at the
  average the percentage of candidates that are visited shown in
  Figure~\ref{fig:dyn_onQueries_visited} gradually decreasing from 12
  to 7\%.  } In terms of throughput, over $900,000$ continuous queries
\indGraph\ is able to serve using a single CPU core an average of $3.2$
million items (tweets) or events (retweets) per minute which is
one order of magnitude more than the number of tweets Twitter actually
receives. Over a total of $100,000$ stored continuous queries the
throughput would go up to $10.9$ million items and events per minute.

\mysubsection{Item/event score weight $\gamma$} In this experiment, we
vary the weight of the dynamic score (aggregating event scores) to the
total score of an item (i.e., the $\gamma$ parameter in
Equation~\ref{eq:totalscore}). Recall that when $\gamma$ is small,
each arriving event has a minimal impact on the total item score and
only a small number of queries will be updated by the corresponding
target items of incoming events. As shown in
Figure~\ref{fig:dyn_onGamma}, \indNaive\ performance improves for
values of $\gamma$ up to 0.2 and then exhibits a constant behavior. On
the contrary, the three other indexes require a higher execution time
for greater $\gamma$ values. In fact, the more $\gamma$ becomes
important, the more the initial static score between queries and items
becomes obsolete. As score changes due to any single event become more
significant, the candidate lists computed on item matching are soon
invalidated given that minimal score of indexed candidate queries has
changed. This leads to an increasing number of false positives and
explains the increase in execution time. For realistic $\gamma$ values
in [0.1, 0.5] the performance difference of any of these three indexes
to their optimal value is of less than 10\%.

\mysubsection{Result size $k$} Figure~\ref{fig:dyn_onK} shows the
performance of the four indexes over different values of the $k$
parameter. Higher $k$ values result in lower values of minimal scores
for the stored queries and consequently increase the number of item
updates received by the queries. As we can see in
Figure~\ref{fig:dyn_onK} all indexes scale linearly on the value of
$k$ and that \indGraph\ and \indSimple\ exhibit a better performance
of about 20\% than the other two indexes when the $k=20$.

\begin{figure}[htbp]
  \centering
  \includegraphics[width=.7\linewidth]{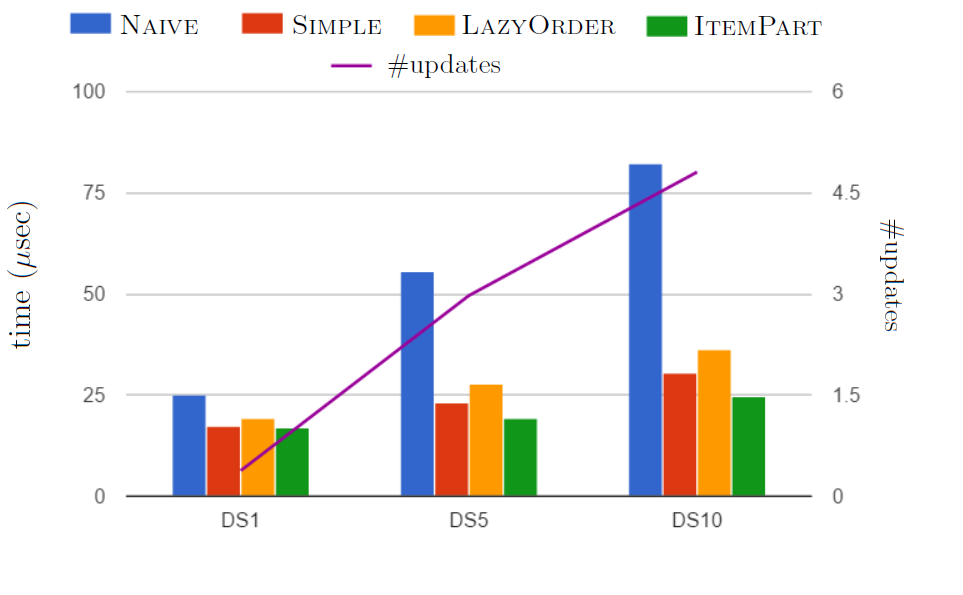}
  \caption{Scalability events per item}
  \label{fig:dyn_onDS}
\end{figure}

\mysubsection{Events per item} In this experiment we measure the
overall performance of our indexes over the three datasets DS1, DS5
and DS10 (see Table~\ref{tab:dyn_datasets}) that feature a different
average number of events each of the items receives: DS1 has an
average of only 1.29 events per item while DS10 has
19.60. Figure~\ref{fig:dyn_onDS} shows per dataset the \emph{average}
execution time for matching a single item or event along with the
average number of updates (\#updates) implied. We can observe that for
all indexes execution time increases proportionally w.r.t. the number
of updates. Additionally, the performance of the \RRTS\ algorithm for
the three indexes (\indSimple, \indLazy\ and \indGraph) is better than
\indNaive\ as the number of events per items increases. Over the DS1
dataset (with the smallest number of events/item), \indGraph, which is
the best performing index, needs 68\% of the time required by
\indNaive, while over DS10 it only requires 29,8\%.

\begin{figure}[htbp]
  \centering
  \includegraphics[width=.7\linewidth]{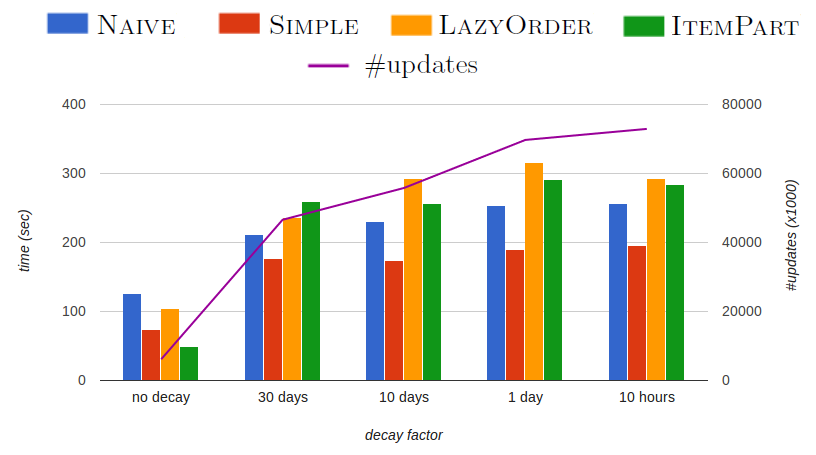}
  \caption{Scalability (linear) decay}
  \label{fig:dyn_onDecay}
\end{figure}

\mysubsection{Decay} In our final experiment in
Figure~\ref{fig:dyn_onDecay}, we measure the performance of all
indexes over linear decay. The horizontal axis shows the time it would
take for a score of 1.0 to decay to 0, i.e., the maximum expiration
time of an ``idle'' item (an item receiving no further events). A
general observation is that as score decay becomes faster, indexes
performance become worse: fast decay leads to a high number of updates
(see number of updates in Figure~\ref{fig:dyn_onDecay}), which in
turn, leads to a higher delay in matching items and events. More
precisely, while \indSimple\ exhibits a 20-30\% overhead over
\indNaive\, the performance of \indLazy\ and \indGraph\ becomes worse
than \indNaive\ even with a small decay factor. This behavior
indicates that the ordering maintained by these two indexes becomes
very dynamic and the stopping condition employed, is not sufficient to
overcome the high maintenance cost. \indSimple\ on the other hand,
does not require any order maintenance and achieves a faster item and
event matching time.

\mysubsection{Conclusions on the experiments}
\label{subsec:dyn_expConclusions}
Our experiments demonstrate that the \indGraph\ and \indSimple\ solutions
outperform the \indNaive\ and \indLazy\ ones over all settings, with \indGraph\ having a slight edge over \indSimple\ of about 5\%. Comparing 
the stopping conditions, \indGraph\ manages to filter with the defined 
stopping condition a big percentage of the candidate lists and on most
cases only had to visit about 10\% of the candidates before having
correctly identified all updates and stopping the algorithm. However,
the heavy data structures maintained for each item (an unsorted set of
sorted lists) only allows \indGraph\ to require from 50 to 35\% the
time used by the \indNaive\ index, which always visits all stored
candidates. Our indexes perform better over highly dynamic
environments with high numbers of events per published item.



%% file: relatedWork.tex
\section{Related work}
\label{sec:relatedWork}


\mysubsection{Real-time search engines}
\longversion{Micro-blog search (e.g., Twitter\footnote{https://twitter.com/search-home}) and Web alert services (e.g., Google Alerts\footnote{www.google.com/alerts}) are typical examples of systems evaluating periodically snapshot queries. When a Twitter user issues a query, a first list of results is immediately returned which regularly gets updated with newer items (tweets). Based on empirical observations on Twitter's website, the top-20 results of user queries are updated every 30 seconds. Google Alerts, can be used over news streams (Google News\footnote{news.google.com}) or over research publications (Google Scholar\footnote{scholar.google.com}). In both cases, a user subscribes giving a text query and is then periodically informed (e.g. once daily) about fresh news articles or scientific publications matching its long-standing query.  

}
Twitter Index (TI) was the first system proposed for real time search~\cite{CLOW11}. It relies on a general form of scoring functions combining linearly textual similarity of items to queries with (static and dynamic) query-independent parameters. A ``\emph{User PageRank}'' authority measure is employed to estimate statically the importance on an item while its dynamic score is computed by counting tweet replies. A rational decay function is applied on the result to capture the freshness of information: $\sTotal(\qu,\ite)/\Delta \tau$, where $\sTotal(\qu,\ite)$ is the initial query-item score and $\Delta \tau$ is the time difference since the publication of the item. TI adopts a partial indexing approach that distinguishes frequent keyword queries from infrequent ones. For any new tweet, if it is relevant to at least one frequent query, it is inserted into an inverted index immediately; otherwise, it is inserted to an append-only log, the content of which is periodically flushed into the inverted index.  However, user queries are evaluated only using the inverted index without looking at the log. The posting lists of this index are ordered on decreasing arrival time of tweets to privilege access to recent tweets even without sufficient consideration of their relevance to queries. 

Deeper insights on how Twitter's Search actually works have been published in~\cite{BGL+12}. To efficiently index all incoming tweets, Earlybird uses compression techniques for tweets' contents and for long posting lists, multi-threaded algorithms and resources allocation optimizations. It relies on an append-only inverted index where each posting list stores tweets sorted by arrival time. This order enables to consider an effective stopping condition: when, during the traversal of the posting list, $k$ items with a similarity score greater than a value $s_{min}$ have been found, the scanning of the posting lists can stop \textit{iff} $s_{min}$ is greater than the maximal value of any other available tweet given that it is published exactly the same time as the one currently being checked. The estimation of this maximal value is based on the score decay function. Several works that followed~\cite{WLXX13,MME+14,LBLT15} proposed improved versions of the inverted index for increasing the accuracy of results returned to real-time search or for supporting various forms of analytic queries without considering continuous query evaluation issues.

\mysubsection{Top-k publish/subscribe on text streams}
\longversion{SIFT~\cite{YGM99} was the first \emph{pull}-based pub/sub system which introduced the continuous query evaluation problem over textual streams. A static threshold per query is used to index only on a subset of terms with high weight. However, this kind of optimization cannot be applied for processing continuous top-$k$ queries where the threshold of each query is the minimal score of its last ($k$-th) result and thus is dynamic. 

}
The Incremental Threshold (IT) algorithm~\cite{MP11} was the first work that introduced continuous top-$k$ queries with sliding information expiration windows to account for information recency. It maintained two inverted indexes: one for queries and the other for items. The latter contains only the most recent items belonging to the current window and its posting lists are sorted by decreasing weight of the terms assigned to represent the items. In the former the posting lists are sorted in decreasing order of a value $\theta_{q,t}$, where $t$ is the term of the posting list and $q$ the continuous query that contains it. IT relies on the Fagin's Threshold Algorithm (TA) to evaluate queries and on an early stopping condition to maintain their result lists. It continuously updates the inverted index on item publication and expiration and thus incurs a high system overhead. This limitation is addressed by the COL-filter algorithm~\cite{HMA10} maintaining only an inverted index of queries with a sorted posting list for each query term. The sorting criteria takes account for the term weight and the minimum query score and allows to define a necessary and sufficient stopping condition for monotonic, homogeneous scoring functions such as cosine or Okapi-25. \cite{HMA10} maintains posting lists where queries are sorted according to the minimum top-k scores and items expire after some fixed time period (window based decay). Minimum top-k scores can change frequently and induce an important resorting overhead. This body of work cannot address continuous top-$k$ queries beyond query-dependent item scores (i.e., content similarity). Efficient stopping conditions over a two-dimensional representation of queries featuring both a query-dependent and query-independent scores (with time decay) have been firstly proposed in~\cite{VAC12}. In this paper, we leverage the item matching algorithm implementing the \IH\ for static scores with time decay and propose several alternative index structures for implementing event matching in the \EH\ (see Figure~\ref{fig:intro_architectureAll}) that accounts for the dynamic score of items due to social feedback.

Finally, \cite{Shraer:2013:TPS:2536336.2536340} considers the problem of annotating  real-time news stories with social posts (tweets). The proposed solution adopts a top-k publish-subscribe approach to accommodate high rate of both pageviews (millions to billions a day) and incoming tweets (more than 100 millions a day). Compared to our setting, stories are considered as continuous queries over a stream of tweets while pageviews are used to dynamically compile the top-k annotations of a viewed page (rather than for scoring the news). The published items (e.g., tweets) do not have a fixed expiration time. Instead, time is a part of the relevance score, which decays as time passes. Older items retire from the top-k only when new items that score higher arrive. This work adapts, optimizes and compares popular families of IR algorithms without addressing dynamic scoring aspects of top-$k$ continuous queries beyond information recency, namely TAAT (term-at-a-time) where document/tweet scores are computed concurrently one term at a time, and DAAT (document-at-a-time) where the score of each document is computed separately. 
\longversion{In particular, three strategies are studied: (a) every new tweet is used as a query for the Story Index and, for every story s, if it is part of the top-k results for s, it is added to the result list Rs (the new tweet is also added to the Tweet Index); (b) for every new story the Tweet Index is queried and the top-k retrieved tweets are used to initialize Rs (the new story is also added to the Story Index); (c) for every page view we simply fetch the top-k set of tweets Rs with no additional processing overhead. }

\mysubsection{Top-k publish/subscribe on spatio-textual streams}
AP (Adaptive spatial-textual
Partition)-Trees~\cite{Wang:2015:AES:2846574.2846648} aim to efficiently serve spatial-keyword subscriptions in location-aware
publish/subscribe applications (e.g.\ registered users interested in local events). They adaptively group user subscriptions using keyword and spatial partitions where the recursive partitioning is guided by a cost model. As users update their positions, subscriptions are continuously \emph{moving}. Furthermore, \cite{Wang:2016:STS:2904483.2904490} investigates the real-time top-k filtering problem over streaming information. The goal is to continuously maintain the top-k most relevant geo-textual messages (e.g.\ geo-tagged tweets) for a large number of spatial-keyword subscriptions simultaneously. Dynamicity in this body of work is more related to the queries rather than to the content and its relevance score. Opposite to our work, user feedback is not considered. 
\cite{Chen:2014:SSE:2733004.2733040} presents SOPS
(Spatial-Keyword Publish/Subscribe System) for efficiently processing spatial-keyword continuous queries. SOPS supports Boolean Range Continuous (BRC) queries (Boolean keyword expressions over a spatial region) and Temporal Spatial-Keyword Top-k Continuous (TaSK) query geo-textual object streams. Finally, \cite{DBLP:conf/icde/ChenCCT15} proposes a temporal publish/subscribe system considering both spatial and keyword
factors. However, the top-k matching semantics in these systems is different from ours (i.e., boolean filtering).

In our work, online user feedback is part of the ranking score of items, which renders existing spatial-keyword solutions inapplicable, while challenges index structures and matching algorithms for top-k continuous query evaluation.



%% file: conclusions.tex
\section{Conclusions}
\label{sec:dyn_discussion}
To support real-time search with user feedback, we have introduced a
new class of continuous top-$k$ queries featuring complex dynamic
scores. The main problem we solved in this context, is the search of
queries that need to update their result lists when new information
items and user events affecting their scores, arrive. To accommodate
high arrival rates of both items and events, we have proposed three
general categories of in-memory indexes and heuristic-based variations
for storing the query candidates and retrieving the result updates
triggered by events. Using an analytic model we have theoretically
proven efficient early stopping conditions avoiding visiting all
candidate queries for each item.  Our experiments validated the good
performance of our optimized matching algorithm. All three indexes for
candidate maintenance (\indSimple, \indLazy\ and \indGraph) achieve a
high throughput of items and events, with \indGraph\ being able to
handle 3.2 million events per minute over 900 thousand stored
continuous queries.  This performance has been achieved using a
centralized single-threaded implementation of our algorithms. The
usage of posting-lists as basic data structures for filtering events
opens a number of opportunities for parallelization.  The main
challenge in this direction concerns the reordering cost of the
ordered solutions (\indLazy\ and \indGraph) with early stopping
conditions. One direction of future work is to consider
workload-oriented parallelization by identifying independent item
clusters (for example, with disjoint query keywords).  Another
direction is to extend the unordered solution \indSimple{}, which is
certainly a good candidate for a first parallel implementation
Finally, we intend to study richer top-$k$ query settings with random
negative feedback scores. The efficiency of our solution is strongly
based on the monotonicity of the scoring function and adding negative
score might need the exploration of completely different
multi-dimensional indexing approaches.

